# Structure from Fleeting Illumination of Faint Spinning Objects in Flight with Application to Single Molecules

*Russell Fung, Valentin Shneerson, Dilano K. Saldin, and Abbas Ourmazd*[*]
*Dept. of Physics*
*University of Wisconsin Milwaukee*
*1900 E. Kenwood Blvd, Milwaukee, WI 53211*

## ABSTRACT

**There are many instances when the structure of a weakly-scattering spinning object in flight must be determined to high resolution. Examples range from comets to nanoparticles and single molecules. The latter two instances are the subject of intense current interest. Substantial progress has recently been made in illuminating spinning single particles in flight with powerful X-ray bursts to determine their structure[1,2], with the ultimate goal of determining the structure of single molecules[3,4,5,6,7]. However, proposals to reconstruct the molecular structure from diffraction "snapshots" of unknown orientation require ~1000x more signal than available from next-generation sources[8]. Using a new approach, we demonstrate the recovery of the structure of a weakly scattering macromolecule at the anticipated next-generation X-ray source intensities. Our work closes a critical gap in determining the structure of single molecules and nanoparticles by X-ray methods, and opens the way to reconstructing the structure of spinning, or randomly-oriented objects at extremely low signal levels. Other potential applications include low-dose electron microscopy, ultra-low-signal tomography of non-stationary objects without orientational information, and the study of heavenly bodies.**

---

[*] Author to whom correspondence should be addressed. Email: ourmazd@uwm.edu.

06/13/2008; 04:24:38 PM

Often, it is of vital interest to determine the structure of a non-stationary object, when neither the position nor the orientation of the object can be controlled. Beyond the obvious example of celestial bodies, examples include a patient's heart during a CT-scan, viruses on a substrate in a microscope, and single molecules illuminated by X-rays. The center-of-mass movement can be "eliminated" by working in diffraction space and the orientation "frozen" by a short snapshot, but the orientation caught in the "snapshot" remains unknown. It was recognized long ago[9], that each diffraction snapshot represents a central section through the diffraction volume in reciprocal space, and, as such, any two snapshots share a "common line." Once the common lines between three snapshots have been identified, the mutual orientations of the three snapshots can be determined. Problems arise, however, when the scattering object is faint. With shot noise present, for example, the common-line approach fails at ~ 10 scattered particles (e.g., photons) per pixel[8]. Faint objects are the rule rather than the exception. A 500kD biological molecule exposed to the intense pulse of a next-generation X-ray source, for instance, scatters only $4 \times 10^{-2}$ photons to a detector pixel at 1.8Å resolution[10,11]. No algorithm capable of reconstructing the structure of such faint objects to high resolution has been demonstrated.

Here, we present a new approach, which exploits the correlations in the entire scattered particle ensemble to recover the structure of a faint object from scattering snapshots of random orientation to high resolution, and demonstrate its power by recovering the structure of a single biological molecule to 1.8Å at scattered fluxes expected from a 500kD molecule in a next-generation X-ray Free Electron Laser. This serves as a general proof of principle, and also closes a key conceptual and algorithmic gap in currently planned "single-molecule" experiments[3,4,5,6,7].

As background for our choice of a demonstration example, we briefly outline the reasons for the intense interest in single-molecule experiments. X-ray crystallography is a powerful tool for determining the structure of macromolecules, but it requires "diffraction quality" crystals. These are often difficult to produce and complicate retrieval of the information of interest: the structure of the molecule. Ideally, one would like to do away with crystals. This has led to proposals to use powerful next-generation X-ray sources, such as X-ray Free Electron Lasers (XFEL's), to determine the structure of individual (i.e., not crystallized) macromolecules and particles. A train of identical objects would be successively exposed to powerful X-ray pulses, and diffraction patterns collected from single objects of unknown orientation. The diffraction patterns would be oriented relative to each other and used to reconstruct the three-dimensional (3-D) diffracted intensity distribution ("the diffraction volume") in reciprocal space. The object structure can then be determined via iterative "phasing algorithms" [12,13,14,15]. In brief, the single-molecule approach represents a "grand challenge" in structure determination.

The primary challenges can be quantified as follows. i) Excluding scattering into information-poor small-angle regions, the number of photons scattered in each "shot" is extremely low. For example, a 500kD molecule exposed to an XFEL beam focused down to 0.1μm scatters ~ $4 \times 10^{-2}$ photons into a detector pixel at 1.8Å resolution in each shot. ii) The incident photon pulse is so intense that the particle explodes within 50fs of pulse arrival, limiting the data collection window to ~ 20fs[16,17]. Beyond experimental difficulties, this also raises questions as to whether the recovered electron density would represent the ground state of the molecule. iii) The approaches proposed for determining the orientation of each diffraction pattern require a mean photon count of ~ 10 per pixel[8], about three orders of magnitude higher than that expected from a 500kD molecule in a 0.1μm diameter XFEL beam. iv) Anticipating this, suggestions have been made first to orientationally classify and average the individual diffraction patterns to boost the signal-to-noise ratio, and then determine the orientation of each class. The per-shot dose needed to "classify" exceeds the available photon flux by two orders of magnitude[8,17]. These difficulties stem from reliance on the very limited information available in one, or a few low-signal diffraction patterns to classify and/or determine orientation.

We begin by outlining the conceptual framework of our approach, first non-mathematically, then in more detail. This is followed by demonstrating the ability to orient simulated low-signal diffraction patterns from a small test molecule, chignolin[18], down to a scattered mean photon count (MPC) of $4 \times 10^{-2}$ per diffraction pattern pixel at 1.8Å, with shot noise - the signal-to-noise level expected for a 500kD molecule. Finally, by recovering the structure of the test molecule from a collection of simulated noisy diffraction patterns of unknown orientation at the MPC of $4 \times 10^{-2}$ per pixel, we show that the orientational accuracy we achieve is sufficient for structure determination to high resolution.

A key challenge is to reconstruct the 3-D diffraction volume for an unknown structure from a collection of extremely low-signal/noise diffraction patterns of unknown orientation. As a loose but illuminating analogy, consider the eroded fragments of an ancient Greek vase recovered in a dig. The vase can be reconstructed from the correlations between the fragments. The most likely shape is obtained when the eroded pieces are maximally correlated with each other. For best results, the correlations considered should not be limited to the shapes of neighboring fragments, but include the elaborate patterns spanning all the fragments. In other words, correlations in the entire data set must be considered simultaneously. This is the basis of our approach: we exploit the correlations in the entire ensemble of diffracted photons to reconstruct the 3-D diffraction volume. An iterative "phasing algorithms" is then used to recover the molecular structure.

In order to describe the conceptual basis of our approach in more detail, a compact nomenclature is needed. Noting that our approach does not rely on any particular data representation, consider the ensemble of scattered photons as a collection of diffraction patterns, each emanating from a random orientation of the object[19]. This allows the discussion to proceed in terms of the familiar diffraction pattern "snapshots." The nomenclature consists of representing each diffraction snapshot by a vector, whose components are the measured intensity values at the pixels of the snapshot. The diffracted photon ensemble is then a matrix consisting of the ensemble of diffraction pattern vectors. The individual pixels in each diffraction pattern span the interval needed to insure optimum information capture[10] ("oversampling" in the sense of [4,12,13,14,15]). (When a gauge satisfies the sampling requirement, it is characterized as the "appropriate sampling" gauge from hereon.)

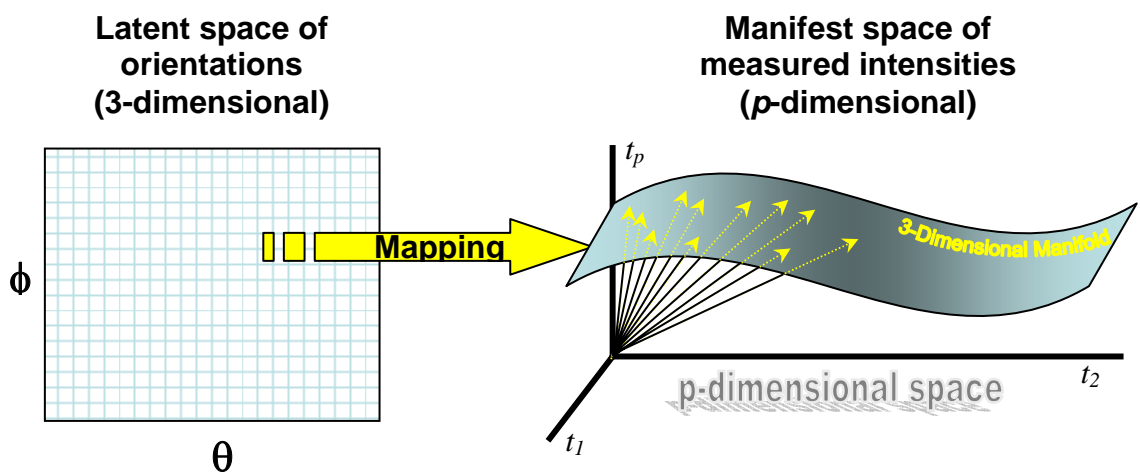

**Fig. 1.** The relationship between the 3-dimensional latent (or hidden) space of orientations and the $p$-dimensional manifest (or accessible) space of measured intensities. The molecular orientation has only three degrees of freedom. As the molecule rotates, the tip of the vector representing its diffraction pattern is confined to a 3-D manifold in the $p$-D manifest space. This manifold is a nonlinear mapping of the space of orientations. The mapping function can be determined by well-known manifold-embedding techniques, and used to relate a particular diffraction pattern to a specific orientation of the molecule.

Now, a molecule in a specific orientation gives rise to a vector in the so-called "manifest space" of measured pixel intensities (Fig. 1). As the molecular orientation is changed in the hidden or "latent space" of orientations, the vector representing the diffraction pattern traces out a path in the $p$-dimensional manifest space of measured intensities. Because the molecule resides in 3-D space, it has only three orientational degrees of freedom. Thus the tip of the vector in the $p$-dimensional intensity space is confined to a 3-D manifold. In order to translate a particular position on the manifold in the manifest intensity space to a specific orientation, i.e., a specific point in the 3-D latent space of orientations, we must determine the mapping between the manifold and the latent space of orientations. This is equivalent to determining the nonlinear function, which maps the latent space to a 3-D manifold in the $p$-dimensional manifest space so as to include all the vector tips. Once this function is known, the position of each vector in the manifest intensity space can be directly related to a point in the latent space of orientations, i.e., to a specific molecular orientation.

The mapping function can be determined by "embedding" a 3-D manifold in the manifest space so as to include all vector tips to within noise, subject to the constraints imposed by the geometry of the latent space. There are a number of manifold embedding techniques. We use Generative Topographic Mapping (GTM), a Bayesian nonlinear factor-analytical approach originally developed for data projection and visualization[20,21,22,23] and neural network[24] applications. This approach determines the maximum likelihood manifold in the manifest space of experimental intensity measurements by fitting the correlations in the diffracted photon ensemble, subject to the constraints imposed by the geometry of the latent space. By determining the nonlinear mapping function, the orientation of each diffraction pattern can be discovered. (For further details, see Supplementary Information.) Through its discrete treatment of the latent and manifest spaces, GTM allows natural classification of similar patterns into orientational classes, and thus noise reduction through averaging. We note, however, that averaging is performed *after* the orientation of each diffraction pattern "snapshot" has been determined. In other words, GTM functions at the actual experimental signal-to-noise level without the need for prior classification and averaging. This is a key attribute.

In order to demonstrate the capabilities of our approach, we have simulated 2-D diffraction patterns of the protein chignolin in random orientations out to a scattering angle corresponding to 1.8Å resolution using an incident photon wavelength of 1Å. Shot noise was incorporated as Poisson statistics. The incident photon intensity was successively reduced so as to produce down to $10^{-2}$ scattered photons per pixel at 1.8Å. Diffraction patterns consisting of (40x40) pixel arrays out to a resolution of 1.8 Å were provided to the program with no information other than the dimensionality of the orientational space. The innermost central pixels were excluded, because, despite their higher photon counts, they contain little orientational information. For each diffraction pattern, up to ~$10^3$ pixels were provided to the algorithm. These stemmed from rectangular strips at the perimeter, annuli excluding the innermost central pixels, or pixels with the highest variance across all diffraction patterns. At incident beam intensities producing a mean photon count of $4 \times 10^{-2}$ per pixel at 1.8Å, the approximately $10^3$ pixels provided to the program typically contained about 100 photons.

We now present our results, starting with the case where the molecule can assume any orientation about one axis. Fig. 2 is a plot of the determined vs. actual orientations for a collection of 3000 diffraction patterns, (a) with no noise (infinite signal), and (b) at a mean photon count of $4 \times 10^{-2}$ per pixel with shot noise. The noise-free case produces a root-mean-square (RMS) orientation error of 1.4° [25]. When the signal is reduced to an MPC of $4 \times 10^{-2}$ per pixel plus shot noise, the RMS orientation error amounts to 3.8° [26]. With our test molecule free to assume any orientation about one axis, the "appropriate sampling angle," the natural scale for the orientational accuracy needed for 1.8Å resolution is 3.2°. We have therefore oriented diffraction patterns to within 1.2 appropriate sampling angles at the mean photon count expected for a single 500kD biological molecule. As shown below, this is ample for structure determination to high resolution.

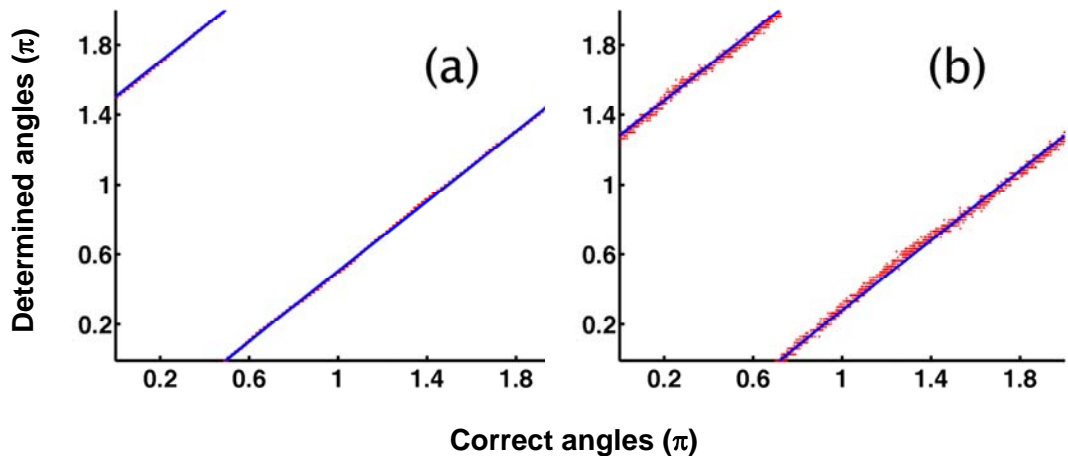

**Fig. 2.** Plot of determined vs. actual orientations (modulo $2\pi$) for 3000 diffraction patterns with: (a) No noise (infinite signal); and (b) Mean photon count of $4 \times 10^{-2}$ per pixel at 1.8Å resolution, with shot noise. The red dots represent actual results, the blue lines best linear fits. The y-intercepts represent unimportant rigid rotations of the molecule.

We next consider the case, where the molecule can assume any orientation in 3-D space. The possible orientations are now represented by points on the surface of the unit 4-sphere[27]. With appropriate sampling for 1.8Å resolution and the test protein free to assume any orientation in 3-D, ~ $10^5$ distinct orientations must be recognized, requiring ~ $10^6$ diffraction patterns. This exceeds our current desktop computational capabilities. We have therefore limited our simulations to random orientations over 30°x30°x30° patches of the surface of the unit 4-sphere. For a set of $10^3$ diffraction patterns with a signal level corresponding to $4 \times 10^{-2}$ photons per pixel at 1.8Å, the root-mean-square error in orientation determination is 5.2°. With the molecule free to assume any orientation in 3-D, the appropriate sampling angle for 1.8Å resolution is 5°. The orientational accuracy achieved is thus 1.04x the appropriate sampling angle. The results demonstrate high orientational accuracy with the molecule able to assume any orientation about one or three axes, albeit over a 30°x30°x30° patch of the surface of the unit 4-sphere in the latter case.

We now show that the orientational accuracy achieved by our approach is sufficient for structure recovery to high resolution. As noted earlier, a full 3-D orientational recovery is beyond our current computing resources. In principle, however, the 3-D molecular structure can be deduced from diffraction patterns obtained when the molecule is free to rotate about a single axis. In practice, the curvature of the Ewald sphere means that only part of the diffraction volume is covered by rotation about one axis, leaving regions devoid of diffraction data. These gaps can be eliminated by allowing the molecule to rotate about each of two orthogonal axes in turn. We have, therefore, used the following procedure to recover the 3-D structure of the test molecule chignolin: i) With the beam along the negative *z*-direction and the molecule free successively to assume any orientation about the *x*- and the *y*-axis, we simulated a total of 72,000 diffraction patterns from random orientations of the molecule. ii) The orientations of the molecule were determined from the diffraction patterns. At an MPC of $4 \times 10^{-2}$ per pixel with shot noise, the RMS orientational error was 1.2x the appropriate sampling angle for 1.8Å resolution. iii) The diffraction patterns belonging to the same orientational classes, each spanning the appropriate sampling angle for 1.8Å resolution were averaged. iv) The data were combined to produce a diffraction volume on a regular Cartesian grid of points in reciprocal space. v) An iterative phasing algorithm[13] was used to recover the structure, which incorporated "charge flipping" of low electron densities[28] and "phase shifting" of weak reflections[29]. Fig. 3 shows the structure recovered to 1.8Å at an MPC of $4 \times 10^{-2}$ per pixel with shot noise. It is clear that the orientational accuracy achieved is ample for high-resolution structure recovery at very low signal levels.

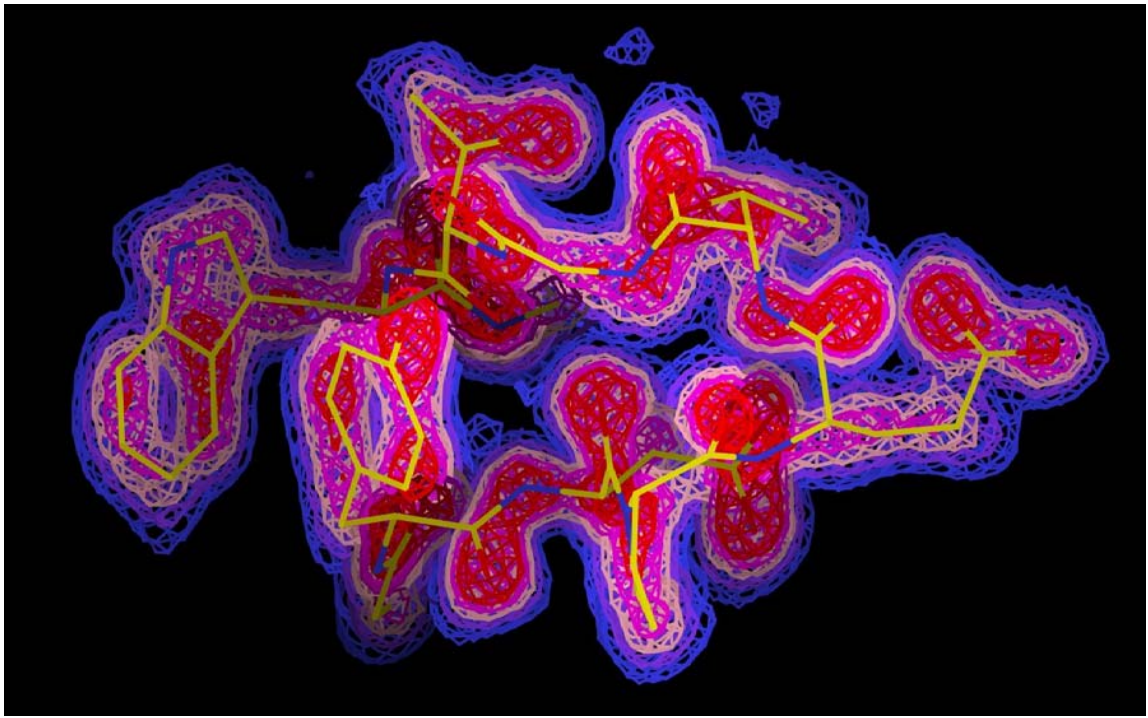

**Fig. 3.** Isosurfaces of electron density of the protein chignolin, recovered from 72,000 diffraction patterns of unknown orientation at a mean photon count of $4 \times 10^{-2}$ per pixel (see text). The molecular model is represented by the stick figure, with C bonds shown in yellow, N in blue, and O in red. The 1, 2, and 3σ electron density contours are shown in blue, pink, and red, respectively, with σ denoting the root-mean-square deviation from the mean electron density. For a 3-D view see: http://www.uwm.edu/~ourmazd/index_files/Page345.htm

We now address the implications of our results for single-molecule structure determination. Using a collection of ~$10^5$ scattered photons, we have so far recovered diffraction pattern orientations at mean photon counts as low as $10^{-2}$ per pixel at 1.8Å. The number of photons scattered to large angle varies as $N^{1/3}$ [8], where $N$ is the number of (non-hydrogen) atoms in the molecule. A 500kD molecule scatters $4 \times 10^{-2}$ photons per pixel to 1.8Å, with heavier molecules producing larger signals. As the signal in effect varies with the cube-root of the molecular weight, an MPC of $10^{-2}$ per pixel, the smallest signal level at which we can at present recover orientation, corresponds to a molecular weight of $500kD \times \left( \frac{10^{-2}}{4 \times 10^{-2}} \right)^3 \sim 8kD$. This represents our current lower limit for the molecular weight, with the upper limit unbounded by intensity considerations.

Available computational resources, however, do set an upper bound. This stems from the increasingly tight orientational accuracy needed for larger objects, with the appropriate sampling angle varying as $(N)^{-1/3}$. A key question, therefore, concerns the computational resources needed to recover the orientations of macromolecules large enough to perform interesting functions. We have carefully characterized the computational requirements of the elementary steps in our approach, and conducted a feasibility study of the resources needed for large molecules and nanoparticles. The results indicate that, with appropriate modifications to the present code and using a computing cluster[30] or a supercomputer, it should be possible to recover the structure of a 500kD molecule to 3Å, a 1MD molecule to 4Å, and a 2MD molecule to 5Å, respectively. We note that this range includes important macromolecules, nanoparticles, and colloids.

The approach we have outlined accurately determines the orientations of diffraction patterns at very low signal levels, thus filling a critical gap in the proposed single-particle experiments. However, these proposals assume a single conformational state for the molecules exposed to X-ray pulses. No means have been suggested to deal with cases where this assumption is not valid. In our approach, if the beam of molecules consists of a number of distinct conformations (or a number of different molecular types), each should produce a different manifold in the manifest intensity space. It should therefore be possible to fit a manifold to each type separately, and determine the structure of a number of distinct molecular (sub)types. By potentially mitigating the need for conformational and/or chemical homogeneity, this would represent a significant advance.

Our algorithm exploits the entire diffracted photon ensemble. So far, we have been able to determine particle orientations at MPC's as low as $10^{-2}$ per pixel using a total of $10^5$ scattered photons. A 500kD particle in an XFEL beam can scatter ~$10^9$ photons to high angle in a few minutes[8]. It may therefore be possible to trade the per-pulse dose against the total number of diffraction patterns recorded, allowing the former to be reduced. If the per-pulse dose can indeed be reduced to below the single-molecule damage threshold, the data collection window increases from the currently anticipated 20fs to the 100ps - 10ns range, depending on the molecular size and rotational energy. Lower per-pulse doses might also bring single-particle structure determination within range of non-FEL

X-ray sources, albeit at reduced resolution.  Two questions then arise: i) What is the lowest practical per-pulse dose needed for structure recovery; and ii) Is this dose below the "acceptable" damage threshold of a single molecule?  These questions highlight important directions for future work.

We now discuss the broader implications of our work.  Our approach reconstructs an object from sections of any shape and dimension with no orientational information.  Potential applications include reconstruction of faint, radiation-sensitive objects by ultralow-dose electron microscopy, diffraction imaging of nanoparticles and colloids, rapid tomography of faint macroscopic objects, whose orientation cannot be controlled, and the study of heavenly bodies.  We note, in addition, that a number of other important classes of problems might be amenable to our approach.  These include, at least in principle, the possibility to recover the distribution of scattering matter (e.g., electron density) directly from an ensemble of diffraction snapshots without "phasing," to parse diffraction patterns of multicrystalline materials into their single-crystal constituents, and to map the energy-wavevector (E, **k**) dispersion surface and hence construct digital energy filters of arbitrary shape.

We acknowledge valuable discussions with M. Schmidt and P. Schwander.  We are grateful to V. Elser for stimulating us to think about general methods for determining orientations, and to D. Starodub for the suggestion to consider the application of our approach to multicrystalline materials.